\documentclass[lettersize,journal]{IEEEtran}
\usepackage{amsmath,amsfonts}
\usepackage{algorithmic}
\usepackage{algorithm}
\usepackage{array}
\usepackage[caption=false,font=normalsize,labelfont=sf,textfont=sf]{subfig}
\usepackage{textcomp}
\usepackage{stfloats}
\usepackage{url}
\usepackage{verbatim}
\usepackage{graphicx}
\usepackage{subcaption}
\usepackage{cite}
\usepackage{xcolor} 
\begin{document}

\title{ Cylindrical RIS-Assisted Low-Complexity Transmission with Differentiated Visible Regions Exploiting Statistical CSI }

\author{Wenjun~Teng, Weicong~Chen, Yiping Zuo, Wankai Tang, and Shi Jin \vspace{-1.0\baselineskip}
\thanks{Wenjun Teng, Weicong Chen, Wankai Tang, and Shi Jin are with the National Mobile Communications Research Laboratory, Southeast University, Nanjing, China (e-mail: twj@seu.edu.cn; cwc@seu.edu.cn; tangwk@seu.edu.cn; jinshi@seu.edu.cn).}
\thanks{ Yiping Zuo is with the School of Computer Science, Nanjing University of Posts and Telecommunications, Nanjing, China (e-mail: zuoyiping@njupt.edu.cn)}
}

\markboth{}%
{Shell \MakeLowercase{\textit{et al.}}: A Sample Article Using IEEEtran.cls for IEEE Journals}


\maketitle

\begin{abstract}
Reconfigurable intelligent surfaces (RIS), recognized as a critical enabler for 6G networks, exhibit unprecedented capabilities in electromagnetic wave manipulation and wireless channel reconfiguration.
By leveraging existing network infrastructure, RIS can cost-effectively create signal hotspots in low-altitude environments, ensuring robust connectivity to support the sustainable development of the low-altitude economy.
However, achieving optimal phase shift design in multi-user scenarios faces two major challenges: the high-dimensional optimization introduced by massive RIS elements, and the persistent coupling of multi-user signals caused by shared RIS reflections.
This paper utilize the visible region of an RIS arranged as the uniform cylindrical array (UCA) to reduce the complexity of phase shift design. 
Under the UCA architecture, RIS elements are categorized into two types: user-specific units and multi-user shared units. 
We then determine the optimal phase shifts by iteratively optimizing the phase shifts of multi-user shared units while directly configuring those of user-specific units based on a derived closed-form solution. 
The proposed approach significantly reduces optimization complexity, which is further corroborated by numerical simulation results demonstrating its substantial impact on both system performance and computational efficiency compared to the conventional RIS with uniform planar array.
\end{abstract}
 

\section{Introduction}
\IEEEPARstart{R}{econfigurable} intelligent surface (RIS) is considered a promising technology for next-generation communication systems\cite{9326394,cui2020information}. 
By adjusting the phase of incident electromagnetic waves to allow their constructive combination at desired locations, RIS improves the quality of transmitted signals, thereby making it particularly valuable for filling coverage gaps and enhancing signal coverage\cite{LYWRISPrinciples,ChannelCustomization,TangPathloss}.
Emerging as a paradigm for transformative urban development, the low-altitude economy (LAE) has garnered considerable attention for its potential to expand the frontiers of human economic activities \cite{10784402,The_Potentialof_Low_Altitude_Airspace,Achieve_Large_scale_Development}. 
This integrated framework leverages the operational synergy of unmanned aerial vehicles (UAVs) within structured airspace, establishing a multidimensional transportation ecosystem.
To support its growth, reliable signal coverage within low-altitude airspace is essential to ensure effective communication between UAVs and base stations (BS) \cite{10693789}. 
RIS technology, in this context, can help create signal hotspots and extend coverage in low-altitude regions, thus improving communication quality while maintaining low cost and power consumption, leveraging existing networks.\setlength{\parskip}{0pt} 

In single-user scenarios, the optimal RIS phase shift can be directly derived via statistical channel state information (CSI) to maximize ergodic spectral efficiency (SE) \cite{han2019large}. However, in multi-user scenarios, obtaining a closed-form solution is challenging, as the RIS serves multiple users simultaneously. 
Considering each user's CSI and other factors further complicates the objective function, necessitating an appropriate optimization method for determining the optimal phase shift.
 In \cite{ren2022performance}, a method based on genetic algorithm is developed for solving the continuous and discrete phase shifts optimization problems with the aim of maximizing the sum achievable security data rate. 
 In \cite{GanX2021Multiuser}, researchers employs alternating direction method of multipliers , fractional programming, and alternating optimization to optimize transmit beamforming and RIS phase shifts, maximizing ergodic sum capacity in multi-user MISO systems under statistical CSI.
Large-scale RIS implementations introduce multidimensional challenges, primarily due to the exponential growth in channel dimensions that intensifies channel estimation complexity \cite{LWChannelEstimation}. 
Simultaneously, the growth of optimization variables in system design is not only intensified by coupling effects among multi-user received signals from shared RIS configurations, but also compounds the complexity of phase-shift optimization through their combined interactions.
These intertwined complexities ultimately demand prohibitive computational resources for optimal solution derivation.\par
 
To address the practical limitations in acquiring instantaneous CSI \cite{CCCE} \cite{guo-arxiv}, we employ statistical CSI for configuring the phase shifts of RIS elements. 
In an effort to further alleviate the computational complexity associated with phase-shift optimization, we propose a novel RIS architecture based on uniform cylindrical array (UCA). 
\textcolor{black}{%
While non-planar RIS configurations have been extensively studied \cite{T3DRIS,ConformalIRS-Empowered,UCA_Low-Complexity}, curved architectures inherently restrict transceiver visibility to element subsets due to their bent geometry. This geometric constraint prevents full-aperture illumination, particularly under large surface curvature.%
}
Under the UCA architecture, RIS elements are categorized as user-specific or multi-user shared based on visible region (VR) analysis for all transceivers.
Aiming at achieving the maximum of system ergodic SE, we investigate phase-shift optimization approaches for different RIS element categories. 
Specifically, we show that the optimal phase shifts for user-specific elements admit closed-form solutions, thereby eliminating the need for iterative optimization like in the case of multi-user shared units and significantly reducing computational complexity.
Finally, comprehensive numerical simulations validate the impact of various RIS physical configurations on both system performance and computational complexity.

\section{SYSTEM MODEL}
\begin{figure}
    \centering
    \includegraphics[width=1.0 \linewidth]{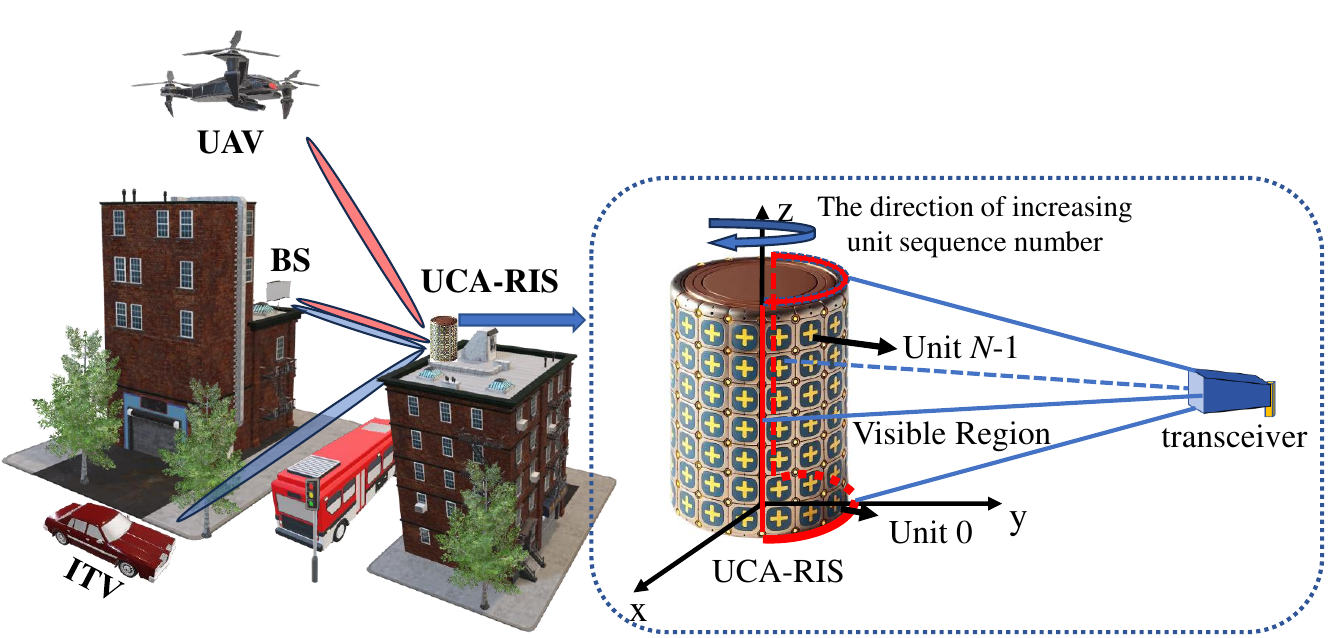}
    \caption{RIS-assisted hybrid low-altitude and terrestrial downlink transmission communication system with explanation of transceiver's VR on RIS}
    \label{fig:sysmodel}
\end{figure}

We consider a RIS-assisted hybrid low-altitude and terrestrial downlink transmission communication system, as shown in Fig.\ref{fig:sysmodel}. 
The BS equipped with an $M$-element uniform linear array (ULA) serves two single-antenna users, that is, a terrestrial intelligent transportation vehicle (ITV) and a Unmanned Aerial Vehicle (UAV).
Due to the obstruction and the downward tilt of the BS antenna, line-of-sight (LoS) paths from the BS to the ITV and to the UAV are blocked, respectively. A RIS configured in a UCA architecture is deployed between the BS and vehicles to assist communications.
Given the cylindrical structure of the UCA-RIS, the BS and vehicles are constrained to specific visible regions (VRs) in the RIS, encompassing no more than half of the RIS's elements. 
This characteristic indicates that only RIS elements within the overlapping VR of the transceiver can built communication link. The RIS is composed of $N_c $ layers, with each layer containing $N_r$ units evenly distributed along a circumference. The total number of RIS units are $N=N_r \times N_c$. 

\subsection{\textit{Channel Model}}  
The channel between the BS and the UAV, composed of the direct non-LoS (NLoS) channel and the RIS-assisted channel, is modeled as:
\vspace{-0.25\baselineskip}\begin{equation}
    {{\mathbf{t}}^H_{\rm{u}}} = {\mathbf{h}}_{\rm{u}}^H{\mathbf{\Phi H}} + {\mathbf{g}}_{\rm{u}}^H \vspace{-0.25\baselineskip},
\end{equation}
where ${\mathbf{h}}_{\rm{u}}^H{\mathbf{\Phi H}\in \mathbb{C}^{1 \times M}}$ is the cascade component and $ \mathbf{g}_\text{u}\in \mathbb{C}^{M \times 1} $ is the direct component component. Adopting Rayleigh fading model, each element of $\mathbf{g}_\text{u}$ is independently and identically distributed  (i.i.d.) and follows a complex Gaussian distribution with zero mean and a variance of $\sigma _{\text{u}}^{2}$ which is determined by the large scale path loss and penetration loss. \par 

In the cascaded link, the channel between the BS to the RIS, $ \mathbf{H}\in \mathbb{C}^{N \times M} $, is modeled as Rician fading. With Rician factors ${\cal K}_{{\text{B}}}$ and large-scale path loss coefficients ${\beta _{{\text{B}}}}$, $\mathbf{H}$ can be expressed by \cite{CCmmWave}:\par \vspace{-1.0\baselineskip}
\begin{equation}
    {\mathbf{H}} = \sqrt {{{{\beta _{{\text{B}}}}}}/({{{\cal K}_{{\text{B}}} + 1}})} ( {\sqrt {{\cal K}_{{\text{B}}}} {\bar{ \mathbf{H}}} + {\tilde{\mathbf {H}}}} ),
    \label{HB} \vspace{-0.5\baselineskip}
\end{equation}

\noindent where $\bar{\mathbf{H}}$ and ${\tilde{\bf H}} $ represent the LoS and NLoS components, respectively. The LoS channel can be expressed as:\vspace{-0.25\baselineskip}
\textcolor{black}{\begin{equation}
\bar {\mathbf{H}}  = ( {{\mathbf{c}}_N^{}( {\varphi _{{\rm{AoA}}}^{{\rm{br}}},\theta _{{\rm{AoA}}}^{{\rm{br}}}} ) \odot {\bf{r}}_N( {\varphi _{{\rm{AoA}}}^{{\rm{br}}}} )} ){\bf{a}}_M^H( {\theta _{{\rm{AoD}}}^{{\rm{br}}}} ),\label{Hlos} \vspace{-0.25\baselineskip}
\end{equation} }
where ${\varphi _{{\rm{AoA}}}^{{\rm{br}}}}$ and ${\theta _{{\rm{AoA}}}^{{\rm{br}}}}$ are the azimuth and elevation angles of arrival (AoA) at the RIS, respectively; ${\theta _{{\rm{AoD}}}^{{\rm{br}}}}$ is the angle of departure (AoD) at the BS; ${{\mathbf{a}}_{M}}( \theta )$, representing the array response vector (ARV) of a ULA with $M$ elements, is given by: \vspace{-1.0\baselineskip}
\begin{equation}
{{\bf{a}}_M}( \theta ) = {[ {1,{e^{j2\pi d\cos \theta /\lambda }}, \ldots ,{e^{j\left( {M - 1} \right)2\pi d\cos \theta /\lambda }}} ]^T},
\label{am} \vspace{-0.25\baselineskip}
\end{equation}
where $d$ denotes the spacing between adjacent antenna elements and $\lambda$ denotes the carrier wave length. In \eqref{Hlos}, the ARV of a UCA with $N$ elements, $\mathbf{c}_{N}^{{}}( \varphi ,\theta  )$, can be expressed as:\vspace{-0.5\baselineskip}
\begin{equation}
    {{\bf{c}}_N}( {\varphi ,\theta } ) = {\bf{a}}_{{N_c}}( \theta  ) \otimes {\rm{ }}{[ {{e^{j\mu \cos ( {\varphi  - {\omega _0}} )}}, \ldots ,{e^{j\mu \cos ( {\varphi  - {\omega _{{N_r} - 1}}} )}}} ]^T},
    \label{cn} \vspace{-0.5\baselineskip}
\end{equation}
where ${\omega _i} = 2\pi i/{N_r}, i=0,..,{N_r-1}$, and $\mu$ is defined as:\vspace{-0.5\baselineskip}
\begin{equation}
    \mu  = \pi d\sin \theta /( {2\lambda \sin ( {\pi /{N_r}} )} ). \vspace{-0.5\baselineskip}
\end{equation}
The array activation vector (AAV), ${\bf{r}}_N (\varphi) $, is introduced to characterize the VR of the BS and vehicles at the RIS. When the BS and vehicles are located in the far-field region of RIS, ${\bf{r}}_N (\varphi) $ can be modeled as:\vspace{-0.5\baselineskip}
\begin{equation}
{{\bf{r}}_N}(\varphi) = {\bar {\bf{r}} _{{N_c}}} \otimes {\tilde {\bf{r}}_{{N_r}}}( \varphi), \vspace{-0.5\baselineskip}
\end{equation}
where ${\bar {\bf{r}} _{{N_c}}} = {\bf{1}} \in {\mathbb{R}}{^{{N_c}\times 1}}$, the $i$-th element of ${\tilde {\bf{r}}_{{N_r}}}( \varphi  )\in {\mathbb{R}}{^{{N_r}\times 1} }$ is determined by:\vspace{-0.5\baselineskip}
\begin{align}
{\tilde r_i} = \{ \begin{array}{l}
1~~{\rm{ if~~cos}}( {\varphi  - ( {i - 1} )2\pi /N} ) \ge 0\\
0~~{\rm{ if~~cos}}( {\varphi  - ( {i - 1} )2\pi /N} ) < 0 .\vspace{-5\baselineskip}
\end{array}  \label{f(a)} 
\end{align}
The Rayleigh fading channel incorporating the AAV ${{\bf{r}}_N^{}( {\varphi _{{\rm{AoA}}}^{{\rm{br}}}} )}$ is used to model the NLoS channel component $\mathbf{\tilde{H}}$, which is expressed as:\vspace{-0.5\baselineskip}
\begin{equation}
 \tilde {\mathbf{H}} = {\tilde {\bf{H}}_\text{A}} \odot ( {{\bf{r}}_N^{}( {\varphi _{{\rm{AoA}}}^{{\rm{br}}}} ){\bf{r}}_M^T} ),\vspace{-0.5\baselineskip}
\end{equation}
where ${{{\bf{r}}}_M} = {\bf{1}} \in \mathbb{R}{^{M \times 1}}$, elements of $\tilde{\mathbf{H}}_\text{A}$ are i.i.d. and follow complex Gaussian distribution with zero mean and unit variance.\par
The channel between RIS and UAV ${\bf{h}}_\text{u} \in \mathbb{C}^{N \times 1}$ is modeled as Rician fading as well. With Rician factor $\mathsf{\mathcal{K}}_\text{u}$ and large-scale path loss coefficient $\beta _\text{u}$, ${\bf{h}}_\text{u}$ can be expressed as:\vspace{-0.25\baselineskip}
\begin{equation}
   {{\mathbf{h}}_\text{u}} = \sqrt {{{\beta _\text{u}}}/({{{\cal K}_\text{u} + 1}})} ( {\sqrt {{\cal K}_\text{u}} {{{\bar{\mathbf h}}}_\text{u}} + {{{\tilde{\mathbf h}}}_\text{u}}} ),\label{hU}\vspace{-0.25\baselineskip}    
\end{equation}
\noindent where $ {\tilde{\mathbf h}}_\text{u} = {\tilde{\mathbf h}}_\text{A} \odot {\bf{r}}_N( {\varphi _{{\rm{Ao}}{{\rm{D}}_\text{u}}}^{{\rm{ru}}}} )$ is the NLoS component, and the LoS component is given by:\vspace{-0.25\baselineskip}
\textcolor{black}{\begin{equation}
{\bar{\mathbf h}}_\text{u}= {\bf{c}}_N^{}( {\varphi _{{\rm{Ao}}{{\rm{D_u}}}}^{{\rm{ru}}},\theta _{{\rm{Ao}}{{\rm{D_u}}}}^{{\rm{ru}}}} ) \odot {\bf{r}}_N( {\varphi _{{\rm{Ao}}{{\rm{D_u}}}}^{{\rm{ru}}}}),\vspace{-0.25\baselineskip}
\end{equation}}
\noindent where \textcolor{black}{${\varphi _{{\rm{Ao}}{{\rm{D_u}}}}^{{\rm{ru}}}}$ and ${\theta _{{\rm{Ao}}{{\rm{D_u}}}}^{{\rm{ru}}}}$} are the azimuth and elevation AoD at the
RIS to UAV, respectively. Similarly to $\tilde{\mathbf{H}_\text{A}}$, elements of ${\tilde{\mathbf h}}_\text{A}\in \mathbb{C}^{N \times 1}$ are i.i.d. and follow complex Gaussian distribution with zero mean and unit variance. \par 
 The channel between  BS and ITV which denoted as ${\bf{h}}_\text{v} \in \mathbb{C}^{N \times 1}$ is similar to the aforementioned channel. And for the sake of simplicity and convenience in subsequent mathematical derivations, 
we use $\mathbf{r}_\text{B}$ to abbreviate  ${{\bf{r}}_N}( {\varphi _{{\rm{AoA}}}^{{\rm{br}}}} )$ and ${{\bf{r}}_\text{u}}$ to abbreviate \textcolor{black}{$ {\bf{r}}_N( {\varphi _{{\rm{Ao}}{{\rm{D_u}}}}^{{\rm{ru}}}} )$}.
\subsection{\textit{System Performance}} 
We assume the BS transmit signal to the UAV and the ITV in different frequency bands. In the absence of user interference, the signals received by the UAV and the ITV remain coupled due to common reflections from the RIS. For instance, the received signal of the UAV is given by: \vspace{-0.5\baselineskip}
\begin{equation}
    {y_\text{u}} = \sqrt {{p_\text{u}}} ( {{\mathbf{h}}_\text{u}^H{\mathbf{\Phi H}} + {\mathbf{g}}_\text{u}^H} ){{\mathbf{f}}_\text{u}}{x_\text{u}} + {n_\text{u}},\vspace{-0.5\baselineskip}
\end{equation}
where $p_\text{u}$ is the transmit power for the UAV, \textcolor{black}{$\mathbf{\Phi} = \mathrm{diag}\{ e^{j\phi_1}, \ldots, e^{j\phi_{N}} \}$ }represents the phase shift matrix,
and $\phi_n \in [0, 2\pi)$ is the phase shift of the $n$-th RIS element. 
The beamforming vector $\mathbf{f}_\text{u} \in \mathbb{C}^{M \times 1}$ satisfies $\| \mathbf{f}_\text{u} \|^2 = 1$, 
the transmitted signal $x_\text{u}$ fulfills $\mathbb{E}\{ |x_\text{u}|^2 \} = 1$, 
and $n_\text{u}$ is additive white Gaussian noise (AWGN) distributed as $\mathcal{CN}(0, \delta_\text{u}^2)$.\par

\section{LOW-COMPLEXITY PHASE SHIFT DESIGN EXPLOITING VR}
In this section, we derive upper bounds of the ergodic SE for the UAV and ITV, and propose an optimal phase shift design scheme for all RIS units using a partial-element gradient descent approach, aiming to reduce computational complexity. 

When ${{\bf{g}}_\text{u}} + {{\bf{d}}_\text{u}}$, where ${\bf{d}}_\text{u}^H = {\bf{h}}_\text{u}^H{\bf{\Phi H}} \in \mathbb{C}{^{1 \times M}}$, is available at the BS, the Maximum ratio transmitting (MRT) is adopted. In this case, the beamforming vector ${\bf{f}}_\text{u}$ can be expressed as:\vspace{-0.5\baselineskip}
\begin{equation}
    {{\bf{f}}_{\rm{u}}} = ( {{{\bf{g}}_{\rm{u}}} + {{\bf{d}}_{\rm{u}}}} )/\| {{{\bf{g}}_{\rm{u}}} + {{\bf{d}}_{\rm{u}}}} \|.\vspace{-0.5\baselineskip}
\end{equation}
By denoting $s_\text{u}={{{p_{\rm{u}}}}}/{{\delta _{\rm{u}}^2}}$, the ergodic SE of the UAV can be expressed as: \vspace{-0.5\baselineskip}
\begin{equation}
C_{\rm{u}} = \mathbb{E}\{ {{{\log }_2}( {1 + s_\text{u}{{\| {{\bf{d}}_{\rm{u}} + {\bf{g}}_{\rm{u}}} \|}^2}} )} \} .\vspace{-0.5\baselineskip}
\end{equation}
\textit{Lemma 1}: The upper bound of the ergodic SE of UAV can be given by:\vspace{-0.25\baselineskip}
\textcolor{black}{\begin{equation}
  C_{\text{u}}^{{\rm{ub}}} = {\log _2}( {1 + {s_{\rm{u}}}( {{\eta _{\rm{u}}}{{\| {\overline {\bf{h}} _\text{u}^H{\bf{\Phi \bar H}}} \|}^2} + {c_{\rm{u}}}} )} ),\vspace{-0.25\baselineskip}
\end{equation}
where ${{\eta _\text{u}}}$ and $c_\text{u}$ are respectively given by:\vspace{-0.25\baselineskip}
\textcolor{black}{}\begin{equation}
{\eta _{\text{u}}} = {{{\beta _{\text{B}}}{\beta _{\text{u}}}{\cal K_{\text{B}}}{\cal K_{\text{u}}}}}/{{( {{\cal K_{\text{B}}} + 1} )/( {{\cal K_{\text{u}}} + 1} )}},\vspace{-0.25\baselineskip}
\end{equation}
\begin{equation}
    c_\text{u}={\chi _\text{u}}M{\bf{r}}_{{\text{B}}}^H{\bf{r}}_\text{u}^{} + \sigma _\text{u}^2M,\vspace{-0.5\baselineskip}
\end{equation}
where ${\chi _{\text{u}}} = {{{\beta _{\text{B}}}{\beta _{\text{u}}}( {{\cal{K}_{\text{B}}} + {\cal{K}_\text{u}} + 1} )}}/{{( {{\cal{K}_{\text{B}}} + 1} )/( {{\cal{K}_{\text{u}}} + 1} )}}$.}
\par
\textit{Proof}: Please refer to Appendix \ref{App:A}. 

In the high signal-to-noise ratio (SNR) regime where ${s_{\rm{u}}}( {{\eta _{\rm{u}}}{{\| {{\bf{\bar h}}_{\rm{u}}^H{\bf{\Phi \bar H}}} \|}^2} + {c_{\rm{u}}}} ) \gg 1$ holds, the upper bound of sum ergodic SE in the RIS-assisted hybrid low-altitude and terrestrial downlink transmission communication system can be approximately given by:\vspace{-0.5\baselineskip}
\begin{equation}
   C_{{\rm{sum}}}^{{\rm{ub}}} = \sum\nolimits_{i \in \{ {\rm{u}},{\rm{v}}\} } {{{\log }_2}( {{s_i}( {{\eta _i}{{\| {{\bf{\bar h}}_i^H{\bf{\Phi \bar H}}} \|}^2} + {c_i}} )} )}.\vspace{-0.5\baselineskip}
   \label{Csum}
\end{equation}
Then, the optimal $\boldsymbol{\Phi}$ that maximize the sum SE can be formulated as: \vspace{-0.5\baselineskip}
\begin{equation}
   \mathop {\max }\limits_{\bf{\Phi }} \sum\nolimits_{i \in \{ {\rm{u}},{\rm{v}}\} } {{{\log }_2}( {{s_i}( {{\eta _i}{{\| {{\bf{\bar h}}_i^H{\bf{\Phi \bar H}}} \|}^2} + {c_i}} )} )}.\vspace{-0.5\baselineskip}   \label{optquestion1}
\end{equation}
It is noteworthy that in problem \eqref{optquestion1}, although the UAV and ITV operate in different frequency bands, optimal phase shifts of multi-user shared units simultaneously affect the received signals of both UAV and ITV, thereby influencing the sum SE of the system. The critical challenge of optimizing RIS phase shifts to maximize the sum SE necessitates rigorous investigation. To resolve this, we develop a hybrid optimization framework that strategically combines partial gradient descent with closed-form solutions based on differentiated VRs for joint RIS phase-shift design, explicitly maximizing the aggregated SE. For user-specific units outside the overlapped VRs, the phase shift design impacts only one of UAV and ITV without affecting the other. Consequently, the optimal phase shifts for these units have closed-form solutions, which are provided in the following lemma:\par
\noindent\textit{Lemma 2}: The optimal phase shift of the user-specific unit $n$ can be given by:\vspace{-0.5\baselineskip}
\textcolor{black}{\begin{equation}
{\varphi _{\rm{n}}}{\rm{ = Arg}}( {( {\sum\nolimits_{k = 1,k \ne n }^N {c_k^{{\rm{bs}}}{{( {c_k^i} )}^*}{e^{j{\varphi _k}}}} } )/c_{n }^{{\rm{bs}}}{{( {c_{n }^i} )}^*}} ) .
\label{bishijie}
\end{equation}}

\noindent where ${c_k^{{\rm{bs}}}}$ and ${c_k^i}$ represent the $k$-th element of 
 ${\bar{\boldsymbol h}}_i$ and $\bar {\boldsymbol{h}}_\text{B}$ respectively. \par 
\noindent\textit{Proof}: Without loss of generality, for the UAV-specific unit $n$, problem (\ref{optquestion1}) can be reformulated as:\vspace{-0.5\baselineskip}
\begin{equation}
    \mathop {\max }\limits_{\bf{\Phi }} {\rm{   }}{\| {{\bf{\bar h}}_{\rm{u}}^H{\bf{\Phi \bar H}}} \|^2}.\vspace{-0.5\baselineskip}
\end{equation}
This implies that the system's sum SE reaches its maximum when the signal regulated by the UAV-specific unit $n$ is perfectly phase-aligned with the combined signal reflected by all other units. In other words, this condition is met when the phase of complex number 
$c_n^{{\rm{bs}}}{( {c_n^i} )^*}{\varphi _{\rm{n}}}$ is identical to that of complex number \textcolor{black}{$\sum\nolimits_{k = 1,k \ne n}^{N } {c_k^{{\rm{bs}}}{{( {c_k^i} )}^*}{e^{j{\varphi _k}}}}$}.\par
The optimal phase shift for multi-user shared units will be computed using a gradient descent-based approach. First introducing  function:\vspace{-0.5\baselineskip}
\begin{equation}
    {f_i}( {\boldsymbol{\varphi }} ) = {\eta _i}{\| {{\bf{\bar h}}_i^H{\bf{\Phi \bar H}}} \|^2} , \label{fphi}\vspace{-0.5\baselineskip}
\end{equation} 
 where ${\boldsymbol \varphi}$ is the diagonal elements of ${\boldsymbol \Phi}$, that is, ${\boldsymbol \Phi}={\text {diag}}({\boldsymbol \varphi})$. Substituting \eqref{Hlos} into \eqref{fphi} and using ${\overline {\bf{h}} _{{\text{B}}}}$ to abbreviate ${{\bf{c}}_N^{}( {\varphi _{{\rm{AoA}}}^{{\rm{br}}},\theta _{{\rm{AoA}}}^{{\rm{br}}}} ) \odot {\bf{r}}_N^{{\rm{bs}}}( {\varphi _{{\rm{AoA}}}^{{\rm{br}}}} )}$, we can further express ${f_i}( {\boldsymbol{\varphi }} )$ as:\vspace{-0.5\baselineskip}
 \begin{equation}
      {f_i}( {\boldsymbol{\varphi }} ) = {\eta _i}{\| {{\bf{\bar h}}_i^H{\bf{\Phi }}{{\overline {\bf{h}} }_{{\text{B}}}}{\bf{a}}_M^H( {\theta _{{\rm{AoD}}}^{{\rm{br}}}} )} \|^2} = {\eta _i}M{\| {{\bf{\bar h}}_i^H{\bf{\Phi }}{{\overline {\bf{h}} }_{{\text{B}}}}} \|^2}.\vspace{-0.5\baselineskip}
 \end{equation}
 We can reformulate the  problem (\ref{optquestion1}) as:\vspace{-0.5\baselineskip}
\begin{equation}
   {\boldsymbol{\Phi }}_{{\rm{opt}}}^{} = \mathop {\min }\limits_{\bf{\Phi }} F( {\boldsymbol{\varphi }} ) = \mathop {\min }\limits_{\bf{\Phi }}  - \prod\limits_{i \in \{\text{U},\text{V}\}} {( {{f_i}( {\boldsymbol{\varphi }} ) + {c_i}} )} .\label{optquestion2}\vspace{-0.5\baselineskip}
\end{equation}
To calculate the partial derivative 
$\partial F( {\boldsymbol{\varphi }} )/\partial {\varphi _n}$ of $F( {\boldsymbol{\varphi }} )$ with respect to ${\varphi _n}$, after introducing the auxiliary vectors ${\boldsymbol{\tau} =\boldsymbol {\varphi}^* \in \mathbb{C}^{N \times 1} }$ and $ {{\boldsymbol{\rho }}_i= {\bf{\bar h}}_i^* \odot {\bf{\bar h}}_{{\text{B}}}\in \mathbb{C}^{ N \times 1}}$,
we can express the function ${f_i}( {\boldsymbol{\varphi }} )$ using auxiliary vectors as follows:\vspace{-0.5\baselineskip}
\textcolor{black}{\begin{equation}
   {f_i}( {\boldsymbol{\varphi }} ) = {\eta _i}M{\boldsymbol{\tau }}_{}^H{{\bf{A}}_i}{\boldsymbol{\tau }},\vspace{-0.25\baselineskip}
\end{equation}}
where ${{\bf{A}}_i} = {{\boldsymbol{\rho }}_i}{\boldsymbol{\rho }}_i^H \in {\mathbb{C}^{N \times N}}$ is Hermitian. $\partial F( {\boldsymbol{\varphi }} )/\partial {\varphi _n}$ can be given by:\vspace{-0.25\baselineskip}
\textcolor{black}{\begin{equation}
    \frac{{\partial {F}( {\boldsymbol{\varphi }} )}}{{\partial {\varphi _n}}} = \frac{{\partial {f_\text{u}}( {\boldsymbol{\varphi }} )}}{{\partial {\varphi _n}}}( {{f_\text{v}}( {\boldsymbol{\varphi }} ) + {c_\text{v}}} ) + ( {{f_\text{u}}( {\boldsymbol{\varphi }} ) + {c_\text{u}}} )\frac{{\partial {f_\text{v}}( {\boldsymbol{\varphi }} )}}{{\partial {\varphi _n}}}. \vspace{-0.25\baselineskip}
    \end{equation}}
And further ${\partial {f_i}( {\boldsymbol{\varphi }} )/\partial {\varphi _n}}$ can be expressed as:\vspace{-0.25\baselineskip}
\textcolor{black}{\begin{equation}
    {{{\partial {f_i}( {\boldsymbol{\varphi }} )}}/{{\partial {\varphi _n}}} = {\eta _i}M{{\partial ( {{\boldsymbol{\tau }}_{}^H{{\bf{A}}_i}{\boldsymbol{\tau }}} )}}/{{\partial {\varphi _n}}}} ,\label{fphi}\vspace{-0.25\baselineskip}
\end{equation}}
where ${{{\partial ( {{\boldsymbol{\tau }}_{}^H{{\bf{A}}_i}{\boldsymbol{\tau }}} )}}/{{\partial {\varphi _n}}}}$ can be given by:\vspace{-0.25\baselineskip}
\textcolor{black}{\begin{equation}
    {{{\partial ( {{\boldsymbol{\tau }}^H{{\bf{A}}_i}{\boldsymbol{\tau }}} )}}/{{\partial {\varphi _n}}}}=p_1+p_2 ,\vspace{-0.25\baselineskip}
\end{equation}}
where \textcolor{black}{$p_1={\sum\nolimits_{{\rm{s = 1,}}s \ne n }^N {{a^i}_{s n  }{e^{j( {{\varphi _{s }} - {\varphi _n} - \pi /2} )}}} }$, $p_2={\sum\nolimits_{{\rm{s = 1,}}s \ne n }^N {{a^i}_{n s}{e^{ - j( {{\varphi _{s }} - {\varphi _n} - \pi /2} )}}} }$}, $a_{mn}^i$ is the element in the $m$-th row and $n$-th column of matrix ${{\bf{A}}_i}$. Since ${{\bf{A}}_i}$ is a Hermitian matrix, it can be proved that $p_1=p_2^*$, and \eqref{fphi} can be further simplified as:\vspace{-0.25\baselineskip}
\begin{equation}
   {{\partial {f_i}( {\bf{\varphi }} )}}/{{\partial {\varphi _n}}} = 2{\eta _i}M{\text{Re}}\{p_1\}.\vspace{-0.25\baselineskip}
\end{equation}

For the shared elements $m$, the phase shift after the $o$-th iteration can be updated according to \eqref{diedai}:\vspace{-0.25\baselineskip}
 \begin{equation}
     \varphi _m^{o + 1} = \varphi _m^o - \varepsilon {{\partial F( {{\boldsymbol{\varphi }}_{}^o} )}}/{{\partial \varphi _m^o}} ,\label{diedai}\vspace{-0.25\baselineskip}
 \end{equation}
\textcolor{black}{where $ \varepsilon  = k \times {10^{ - ( {{{\log }_2}N+t} )}}$ is the iteration step size. In the subsequent simulations, $k$ is set to 1 and $t$ is set to -4, which are parameters used for finer adjustment of the iteration step size. The objective function value increases logarithmically with the number of RIS elements $N$. Hence, to accelerate convergence for large $N$, larger iteration step sizes can be employed by decreasing $t$ and increasing $k$. }Notably, not all phase shifts require gradient-based optimization because of the VR, implying that the UCA configuration of the RIS brings in lower computational complexity for optimization. Moreover, after each iteration, the phase shifts of the user-specific units can be configured according to \eqref{bishijie}  to achieve the current local optimum and enhance the gradient, thus accelerating the iterative computation process. For example, in a scenario where the azimuth AoA for two vehicles are symmetric with respect to the BS, the average computational complexity of using a UCA architecture for phase shift design based on the gradient descent algorithm is  less than half that of using a UPA architecture. Furthermore, when the azimuth AoA exceeds 90 degrees, the computational complexity reduces to $\mathcal{O}(1)$.
\section{NUMERICAL RESULTS}
In this subsection, we primarily compare the system performance and the complexity of phase shift design corresponding to RIS configured in UCA mode and UPA mode. Unless otherwise stated, $\sigma_\text{u}^2 = \sigma_\text{v}^2 = -110$, transmit power $p_\text{u}^2 = p_\text{v}^2 = 30\ \mathrm{dBm}$, noise variance $\delta_\text{u}^2 = \delta_\text{v}^2 = -107\ \mathrm{dBm}$, Rician factors $\mathcal{K}_\text{B} = \mathcal{K}_\text{u} = \mathcal{K}_\text{v} = 13\ \mathrm{dB}$, large-scale path loss coefficients $\beta_\text{B} = k_0 d_\text{B}^{-\alpha_\text{B}}$, $\beta_\text{u} = k_0 d_\text{u}^{-\alpha_\text{u}}$, $\beta_\text{v} = k_0 d_\text{v}^{-\alpha_\text{v}}$, where $k_0 $ is set to $10^{-3}$, with path loss exponents $\alpha_\text{B} = 2.0$, $\alpha_\text{u} = 3.0$, and $\alpha_\text{v} = 2.0$. The corresponding communication distances are configured as $d_\text{B} = 80\ \mathrm{m}$, $d_\text{u} = 80\ \mathrm{m}$, and $d_\text{v} = 400\ \mathrm{m}$. Parameters of BS and UCA-RIS units are set as follows: $M=16$, $N_c=4$, $N_r=64$, $\varphi_\text{AoA}^\text{br} = 0^\circ$, $\theta_\text{AoA}^\text{br} = 90^\circ$. \textcolor{black}{To ensure a fair comparison, we maintain a constant number of RIS units within the BS's VR across different architectures, so the number of UPA-RIS elements is set to half of that in the UCA-RIS during the simulation.}\par

\begin{figure}
    \centering 
    \includegraphics[width=1\linewidth]{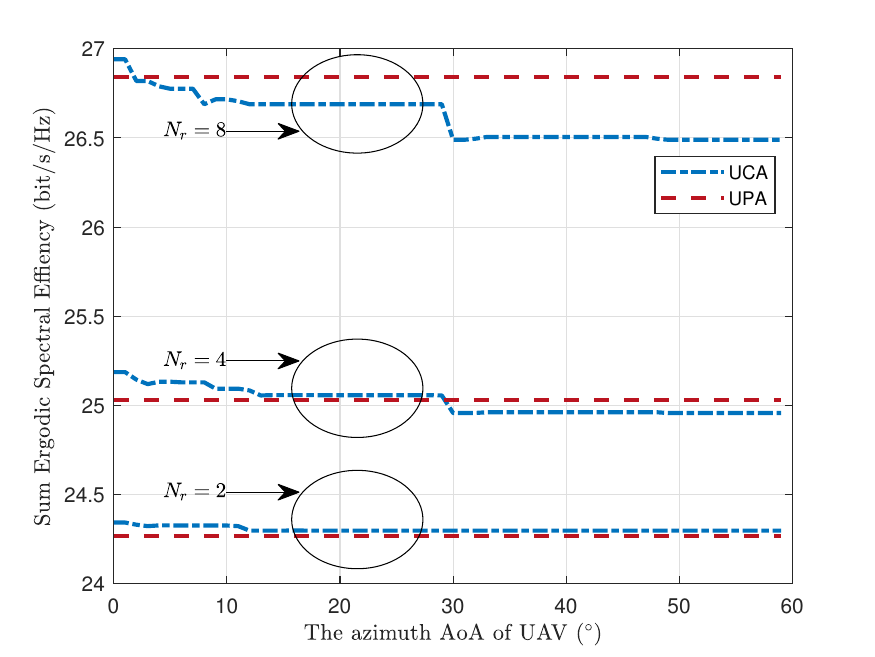} 
    \caption{System Performance vs location of UAV}\vspace{-1\baselineskip}
    \label{fig:conclusion1}
\end{figure}

Fig. \ref{fig:conclusion1} illustrates the relationship between system performance and the azimuth AoD of UAV when the two vehicles are symmetrically distributed with respect to the BS within the region. 
It is worth noting that when the azimuth AoD of UAV varies within a certain range near the BS, the UCA architecture for the RIS achieves higher system performance compared to the UPA architecture. 
This observation suggests that partitioning the VR of different users on the RIS, such that each element serves only one user without interfering with others, can further enhance system performance. 
The performance enhancement incurs a fundamental tradeoff through diminished utilization of effective reflecting RIS elements. 
Simulation results reveal that with increasing UAV azimuth AoD, particularly for more layers, the UCA configuration becomes inferior to UPA configuration. 
This demonstrates that while single-user-exclusive operation prevents inter-user interference, its advantages cannot compensate for the SE degradation caused by reduced effective element utilization, ultimately diminishing overall system performance. 
The new equilibrium point shifts towards a smaller azimuth AoD of UAV as the number of RIS layers increases.
Furthermore, the UCA architecture demonstrates pronounced angular sensitivity in sum SE compared to UPA architecture, owing to dynamic transceiver VR overlap variations induced by UAV azimuth AoD changes. Unlike the invariant VR overlap in UPA architecture, the sum SE decreases as the overlapping VR diminishes in UCA architecture. 
\textcolor{black}{When the RIS adopts a UPA configuration, the overlapping VR between the UAV and the BS corresponds to the entire RIS array. Therefore, for a UAV at the same position, the difference in the number of units within the overlapping VR between a UCA-RIS and a UPA-RIS increases as the number of layers increases. In this case, the gain from configuring phase shifts for user-specific units must counterbalance the reduction in array gain caused by the larger difference in the number of units in the overlapping region. As a result, the equilibrium point shifts toward a smaller UAV azimuth AoD, because the difference in the number of VR units decreases as the overlapping region shrinks.}
\par
\begin{figure}
    \centering 
    \includegraphics[width=0.85\linewidth]{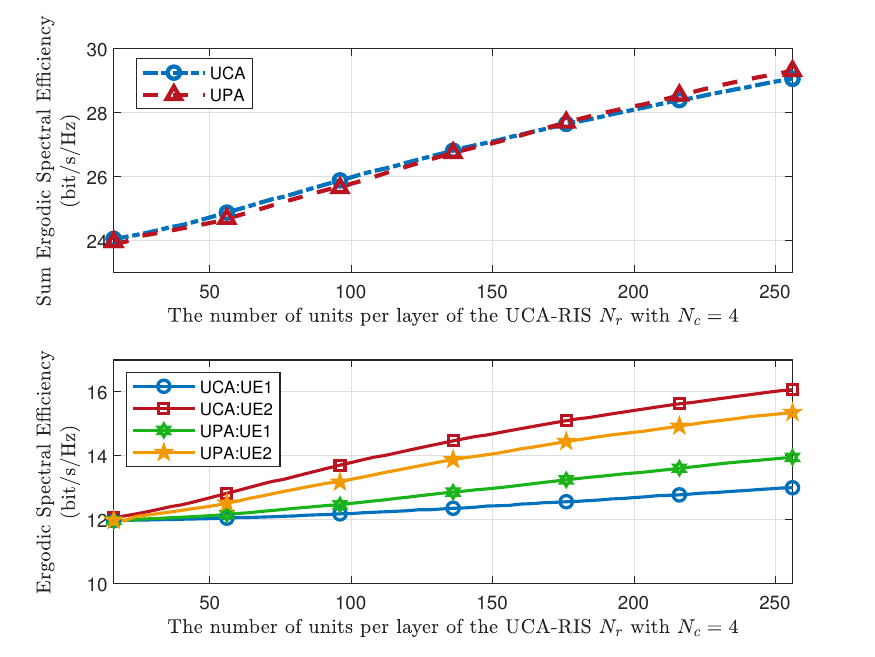} 
    \caption{System Performance vs number of RIS units}\vspace{-2\baselineskip}
    \label{fig:conclusion2}
\end{figure}
Figure \ref{fig:conclusion2} illustrates the relationship between the sum ergodic SE, individual vehicle's ergodic SE and the number of RIS units per layer $N_r$, when the positions of the two vehicles are fixed. 
The simulation results show that as $N_r$ increases, both the sum ergodic SE and individual vehicle's ergodic SE improve. However, when the RIS adopts the UCA architecture, the growth rate of the sum ergodic SE with respect to $N_r$, becomes slower. 
This is because, with fixed vehicle positions, an increase in $N_r$ not only enhances the RIS array gain but also leads to a higher number of units in non-overlapping VR of the transceiver, resulting in a slower rate of performance improvement compared to the UPA architecture. 
Nevertheless, when $N_r$ is large, the UCA architecture offers the advantage of achieving nearly equivalent performance to the UPA architecture while reducing computational complexity. 
Furthermore, when the RIS employs the UCA configuration, the final optimization results reveal a larger SE disparity between the two vehicles. 
This occurs because adjusting the phase shifts of user-specific units during iteration amplifies the SE gap between the two users, causing the user with higher SE to contribute more significantly to the gradient computation. Consequently, the optimized RIS exhibits stronger beam directivity, meaning the beam aligns more precisely with the user having better link quality.
\par
\begin{figure}
    \centering 
    \includegraphics[width=1\linewidth]{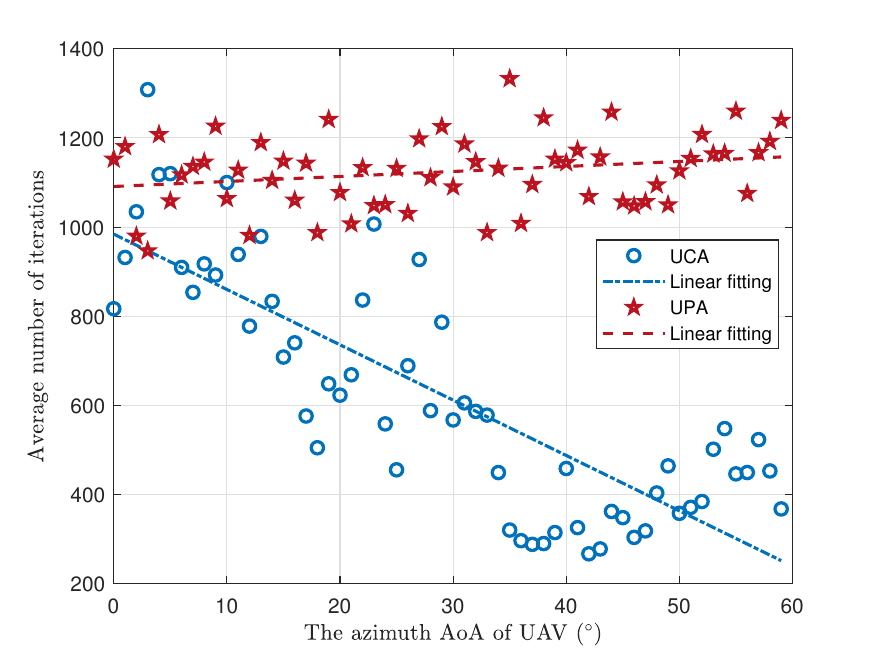} 
    \caption{Optimize complexity vs location of UAV}
    \label{fig:conclusion3}\vspace{-1.5\baselineskip}
\end{figure}
Figure \ref{fig:conclusion3} compares the average number of optimization iterations for RIS under two different architectures, with each scenario's results averaged over 100 optimization trials. The numerical simulation results indicate that under the UPA architecture, the average number of iterations remains nearly constant. 
In contrast, with the UCA architecture, the average number of iterations decreases as the azimuth AoD of UAV increases. 
This is because the number of user-specific units increases with the azimuth AoD of UAV, and these elements do not require gradient calculations.
Instead, their phase shifts can be directly computed using \eqref{bishijie}, thus reducing the optimization complexity.
Moreover, when employing the UCA architecture for iterative optimization, the algorithm not only updates the current phase shift vector based on gradient calculations but also adjusts the phase shifts of user-specific units according to their computed values.
This ensures that the current solution is locally optimal, which accelerates the algorithm’s convergence. 
Under the UPA architecture, however, numerical results demonstrate that the number of iterations can be up to four times higher than those required under the UCA architecture.
\section{Conclusion}
\textcolor{black}{To reduce the computational complexity of phase shift optimization for RIS-assisted multi-user scenarios, this study introduced a novel framework that explicitly partitions the VRs of users on the RIS. Specifically, by categorizing RIS units into user-specific and multi-user shared subsets, the proposed method allows the phase shifts of user-specific elements to be determined via closed-form expressions, thus enhancing algorithmic efficiency without sacrificing performance. In addition, a comprehensive performance evaluation was carried out to investigate the impact of different RIS physical architectures. Numerical results demonstrated that the proposed UCA-RIS not only achieves performance comparable to—or even surpassing—that of conventional UPA-RIS, but also offers substantial reductions in computational complexity, particularly when the number of RIS layers is moderate. These findings highlight the practicality and scalability of the proposed UCA-based design for future RIS-assisted multi-user communication systems.}

{\appendix[Proof of  \textit{Lemma 1}]
\label{App:A}
According to Jensen's inequality, the upper bound of the ergodic SE of UAV can be expressed as:\vspace{-0.5\baselineskip}
\begin{equation}
    C_{\rm{u}}^{{\rm{ub}}} = {\log _2}( {1 + s_\text{u}\mathbb{E}\{ {{{\| {{\bf{d}}_{\rm{u}} + {\bf{g}}_{\rm{u}}} \|}^2}} \}} ).
    \label{CubappA}\vspace{-0.5\baselineskip}
\end{equation}
Since ${{\bf{d}}_{\rm{u}}}$ and ${{\bf{g}}_{\rm{u}}}$ are mutually independent, we can obtain:\vspace{-0.5\baselineskip}
\begin{equation} 
    {\mathbb{E}\{ {{{\| {{{\bf{d}}_{\rm{u}}} + {{\bf{g}}_{\rm{u}}}} \|}^2}} \}} =\mathbb{E}\{ {{{\| {{{\bf{d}}_{\rm{u}}}} \|}^2}} \} + \mathbb{E}\{ {{{\| {{{\bf{g}}_{\rm{u}}}} \|}^2}} \} .\label{indpendent}\vspace{-0.5\baselineskip}
\end{equation}
By following a similar derivation process as in \cite{han2019large}, we can obtain
${\mathbb{E}}\{ {{{\| {{{\bf{g}}_{\rm{u}}}} \|}^2}} \} = \sigma _{\rm{u}}^2M$ and $\mathbb{E}{\{ {{{\| {{{\bf{d}}_{\rm{u}}}} \|}^2}} \}}$ can be given as follow:\vspace{-0.5\baselineskip}
\begin{equation}
\mathbb{E}{\{ {{{\| {{{\bf{d}}_{\rm{u}}}} \|}^2}} \} = {\eta _{\rm{u}}}(\mathbb{E}\{ {{{\| {\overline {\bf{H}} {\bf{\Phi }}{{\overline {\bf{h}} }_{\rm{u}}}} \|}^2}} \} + x_1+x_2+x_3 ) }, 
\end{equation}
\textcolor{black}{and $x_1$ is formulated as.:\vspace{-0.5\baselineskip}
\begin{equation}
     x_1={\mathbb{E}}\{ {{{\| {\lambda \widetilde {\bf{H}}{\bf{\Phi }}{{\widetilde {\bf{h}}}_{\rm{u}}}} \|}^2}} \} = {\lambda ^2}M{\bf{r}}_{\text{B}}^H{{\bf{r}}_{\rm{u}}},
    \label{NLOS3}\vspace{-0.5\baselineskip}
\end{equation}
$x_2$ is given by:\vspace{-0.5\baselineskip}
\begin{equation}
   x_2={\mathbb{E}}\{ {{{\| {\lambda \sqrt {{\cal{K}_{\rm{u}}}} \widetilde {\bf{H}}{\bf{\Phi }}{{\overline {\bf{h}} }_{\rm{u}}}} \|}^2}} \} = {\lambda ^2}{\cal{K}_{\rm{u}}}M\mathbf{r}_\text{B}^H{{\bf{r}}_{\rm{u}}},
   \label{NLOS2}\vspace{-0.5\baselineskip}
\end{equation}
$x_3$ is expressed as:\vspace{-0.5\baselineskip}
\begin{equation}
    x_3={\rm{E}}\{ {{{\| {\lambda \sqrt {{\cal{K}_{\rm{B}}}} \overline {\bf{H}} {\bf{\Phi }}{{\widetilde {\bf{h}}}_{\rm{u}}}} \|}^2}} \} = {\lambda ^2}{\cal{K}_{\rm{B}}}M\mathbf{r}_\text{B}^H{{\bf{r}}_{\rm{u}}}.
    \label{NLOS1}\vspace{-0.25\baselineskip}
\end{equation}}
where \textcolor{black}{$\lambda  = \sqrt {\frac{{{\beta _{\rm{B}}}{\beta _{\rm{u}}}}}{{( {{\cal{K}_{\rm{B}}} + 1} )( {{\cal{K}_{\rm{u}}} + 1} )}}}$and the proof of \textit{Lemma 1} is completed.}}

\bibliographystyle{ieeetr}
\bibliography{Ref}

\end{document}